# Holographic cosmology with logarithmic equation of state based on a new generalized entropy

*I. Brevik[1]*

[1]*Department of Energy and Process Engineering,*

*Norwegian University of Science and Technology, N-7491 Trondheim, Norway*

*A. V. Timoshkin[2,3]*

[2]*Tomsk State Pedagogical University (TSPU), 634041 Tomsk, Russia*

[3]*Lab. for Theor. Cosmology, International Centre of Gravity and Cosmos,*

*Tomsk State University of Control Systems and Radio electronics*

*(TUSUR), 634050 Tomsk, Russia*

*iver.h.brevik@ntnu.no*

*alex.timosh@rambler.ru*

The cosmological application of a dark viscous fluid in a spatially flat Friedmann-Robertson-Walker (FRW) universe with a modified logarithmic equation of state (EoS), being an analogue of the equation associated with the isotropic deformation of crystalline solids, is explored. This formalism represents a class of the so-called logotropic fluids, and allows explaining an accelerating late-time universe. Our research is based on a new generalized entropy function proposed by Nojiri-Odintsov-Faraoni [1]. We consider entropic cosmology and describe the evolution of the late time universe in the presence of dark matter coupled with viscous dark energy. In order to obtain a more detailed picture of its evolution, we add in our model a coupling of the log-corrected power-law fluid with dark matter, and study various interacting forms between them. We solve the system of dynamic equations for viscous dark fluid coupled with dark matter and obtain expressions for the log-corrected power-law dark energy density, and the energy density for dark matter.

The research results are presented in holographic form in terms of generalized holographic cut-offs introduced by Nojiri and Odintsov [2, 3]. In the asymptotic case an expression for the Hubble function, and the energy density for dark matter in the entropic cosmology, are obtained.

## I. Introduction

. In 1998, using astronomical observational data obtained independently in the laboratories of A. Riess and S. Perlmutter [4, 5], it was proven that the expanded Universe is accelerating. Explaining the physical origin of cosmic acceleration in the early and late Universe is a fundamental problem in theoretical physics of the 21st century.

According to astronomical observations, up to 73% of the energy density of the Universe is currently made up of a component known as dark energy. The remaining about 27% is cold dark matter and only 4% is ordinary baryonic matter, which is concentrated in galaxies and their clusters. Dark energy does not interact with ordinary matter; it can be interpreted as a cosmological constant (vacuum energy density).

One of the most interesting models describing dark energy is the ideal fluid model with an unusual equation of state. This fluid must have negative pressure and negative entropy.

A dark fluid can be characterized by the equation of state parameter $\omega = \frac{p}{\rho}$, where $\rho$ is the dark energy density and $p$ is the dark energy pressure. According to recent astronomical observations, the value is equal to $\omega = -1.04^{+0.09}_{-0.10}$ [6]. The most general models of dark fluid can be described using an inhomogeneous equation of state. It is precisely such models and their consequences that will be studied in this work.

The theory predicts many interesting ways for the evolution of the Universe in the future, including Big Rip cosmology [7, 8], Little Rip [9-17], Pseudo Rip [18], and Quasi Rip [19, 20] cosmological models. In the early Universe there is actually the possibility of realizing not only inflationary acceleration, but also a cosmological model of matter with a bounce [21- 24].

If the thermodynamic parameter in the (EoS) is $(\omega < -1)$, the phantom dark energy can lead to a Big Rip (BR) future singularity (for a general classification of finite-time future singularities see Nojiri-Odintsov-Tsujikava, [25]).

An important topic in modern cosmology is the study of the influence of the interaction of dark energy with dark matter in relation to accelerated cosmic expansion [26]. This can be done by including the coupling of dark energy with dark matter [27, 28].

The dissipative properties of the Universe interacting with dark matter can be described in terms of entropy cosmology. A new generalized entropy function, depending on four parameters, was introduced by Nojiri-Odintsov-Faraoni in [1, 29]. Within the framework of another model of entropy cosmology, proposed Odintsov-Paul in [30], the nonsingular behavior of the entropy function allows us to describe the bounce scenario when the Universe passes through the zero value of the Hubble function $H = 0$ at the moment of the bounce. In general entropy cosmology, the Friedmann equations follow from the fundamental laws of thermodynamics, since the entropy function generates the energy density and pressure in these equations. The new entropy function generalizes the properties of previously proposed Tsallis [31], Barrow [32], Renyi [33], Kaniadakis [34], Sharma-Mittal [35], and Loop Quantum Gravity [36] entropies.

We will present the results of our research in a holographic form. This mathematical formalism is based on the application of the holographic principle. It is understood that all information about the parameters of the system can be described in the form of a hologram associated with the surface area of outer space [37]. A generalized model of holographic dark energy (HDE) and a combination of phantom inflation with phantom acceleration was proposed in [2, 3]. It has been shown that the Nojiri - Odintsov model generalizes all known HDE models [38-40]. Astronomical observations confirm the holographic theory of the description of the Universe [41-45].

We will explore in entropy cosmology the dynamics of the evolution of the late Universe using a logarithmically corrected power equation of state taking into account the bulk viscosity of the dark fluid. A spatially flat Friedman-Robertson-Walker metric, homogeneous and isotropic on a large scale is considered.

The results of astronomical observations confirm that the standard theoretical model (CDM) of cold dark matter agrees well with experimental data on a large (cosmological) scale, but is inconsistent on a small (galactic) scale. The problem that arises is related to the assumption that dark matter has no pressure. It is shown that the description of the evolution of the late Universe on cosmic small scales can be formulated on the basis of the formalism of the logarithmically corrected power-law equation of state of matter within the framework of the Debye approximation [46, 47]. In this model, the fluid pressure is modeled using an empirical formula for the pressure of crystalline solids deformed under isotropic stress [12]. In order for the Universe to accelerate under the influence of cosmic expansion, it is necessary that the fluid pressure described by the equation of state be negative [49-51]. However, negative pressure in the logarithmically corrected power-law dark fluid model becomes dominant when the volume of the universe exceeds a certain value. This scenario corresponds to the approaches of the logo tropic dark energy (LDE) model [52, 53]. It is known that there are regimes in which the general properties of logarithmically adjusted power-law dark energy and logo tropic dark energy become equivalent.

We also note other works on this topic. Cosmological applications of the theory of the single component viscous fluid having a logarithmic equation of state have been studied in [54]. A more detailed picture in the log-corrected power-law viscous two-component fluid model containing a coupling with dark matter was studied in [55]. A nonlinear logarithmic interaction was studied in [56].

To describe the acceleration of the late Universe in more detail, a two-component coupled fluid model is used. As the second component of the fluid, dark matter without pressure is considered, weakly interacting with the logarithmically corrected power-law fluid. We will analyze various types of interactions between dark energy and dark matter, considering various options for bulk viscosity and obtain solutions to the gravitational equations of motion.

The motivation of the present investigation is to study the nature of the evolution of the late universe based on the thermodynamic approach in entropy cosmology.

## II. Viscous fluid with logarithmic equation of state

Let us study a dark energy scenario based on a modified logarithmically corrected power-law equation of state for a dark fluid. The (EoS) has the form [57, 58]:

$$p = A\left(\frac{\rho}{\rho_*}\right)^{-l} \ln\left(\frac{\rho}{\rho_*}\right), \tag{1}$$

where $\rho_*$ is a reference density which will be identified with the Planck density [16]: $\rho_p = c^5/\hbar G^2 \approx 5.16 \times 10^{99}\, gr/m^3$. Hear the parameter $A > 0$ may be taken to represent the logotropic temperature, while $l = -\frac{1}{6} - \gamma_G$ ($\gamma_G$ is the no dimensional Gruneisen parameter). The parameter $A$ is determined by the relation $A = B\rho_\Lambda c^2$ [53]. The dimensionless logotropic temperature is approximately given by $B \approx 1/\ln(\rho_P/\rho_\Lambda) \approx 1/[123\ln(10)]$, where $\rho_P$ and $\rho_\Lambda$ are the Planck density and the cosmological density correspondingly. This follows a new interpretation of the famous number $123 \Box \log(\rho_P/\rho_\Lambda)$ as an inverse logotropic temperature.

For $l = 0$ we obtain the equation of state for the logo tropic cosmological model [59].

Let us rewrite the equation (1) following the notation of the LDE model. For this purpose we express the volume in terms of mass density, using the relation [57, 58]:

$$p(V) = -\beta\left(\frac{V}{V_0}\right)^{-\frac{1}{6}-\gamma_G} \ln\left(\frac{V}{V_0}\right), \tag{2}$$

where $V_0$ is a volume which presents a barrier among the different signs of the pressure $p$, and $\beta$ is the bulk modulus at $V_0$. The bulk modulus shows how much the volume change under the action of external forces. The parameter $\gamma_G$ in the homogeneous and isotropic universe is a free parameter of the theory. When $V < V_0$ the pressure is positive for the positive bulk modulus and it is negative in the case of an inequality of the opposite sign.

If the pressure of the dark fluid satisfies the equation (2), then in order to ensure the cosmic acceleration of the universe the volume must overcome the barrier $V \approx V_0$. There are three different regimes of the behavior of the pressure (2) [57]:

1. The time before passing the $V_0$ barrier, when $V < V_0$. Then the pressure is positive and the universe is decelerating. This case corresponds to the case of the pressure less matter in the $\Lambda CDM$ model.

2. The time of equivalence between volumes, when $V = V_0$. That is the transition time from deceleration of the universe to acceleration one.

3. The time after passing the $V_0$ barrier, when $V > V_0$. Then the pressure is negative and the fluid starts to accelerate the universe.

Thus, in the log-corrected power-law model the dynamical evolution of the universe is described by a single fluid, which accelerates the universe, when its volume passes the barrier $V = V_0$. This allows us to apply this model to the description of the late universe.

We will study the dynamical evolution of the universe using a fluid described by the log-corrected power-law equation of state (1). We consider a homogeneous and isotropic flat universe, taking into account the viscosity property of the fluid. To do this, we modify equation (1), adding the term with viscosity [60]:

$$\zeta(H,t) = \xi_1(t)(3H)^n, \tag{3}$$

where $\zeta(H,t)$ is the bulk viscosity, which depends on the parameter Hubble $H$ and on the time $t$. From a thermodynamic point of view it follows, that $\zeta(H,t) > 0$. Thus, taking into account the above, the (EoS) of a logarithmically corrected power-law fluid will take the form:

$$p = A\left(\frac{\rho}{\rho_*}\right)^{-l} \ln\left(\frac{\rho}{\rho_*}\right) - 3H\zeta(H,t). \tag{4}$$

This (EoS) is the analog of the (EoS) for dark energy and is a particular case of the generalized (EoS).

For the review of viscous cosmology, see [61].

**III. New generalized entropy**

The generalized entropy function $S_g$ was proposed in [1],

$$S_g(\alpha_+, \alpha_-, \beta, \gamma) = \frac{1}{\gamma}\left[\left(1 + \frac{\alpha_+}{\beta}S\right)^\beta - \left(1 + \frac{\alpha_-}{\beta}S\right)^{-\beta}\right], \tag{5}$$

where $\alpha_+, \alpha_-, \beta, \gamma$ are positive parameters and $S$ is the Bekenstein-Hawking entropy. The latter is the thermodynamical entropy of a black hole [62, 63],

$$S = \frac{C}{4G}, \qquad (6)$$

where $C = 4\pi r_h^2$ is the horizon area and $r_h$ is the horizon radius. The Bekenstein-Hawking entropy describes the radiation from a black hole.

Various forms of entropy other than Bekenstein-Hawking entropy have been proposed in [31-36]. These entropies have common properties: in the appropriate limit they reduce to the Bekenstein-Hawking entropy and are monotonically increasing functions of the Bekenstein-Hawking entropy variable. It has been proven in [1] that the generalized entropy function $S_g$ reduces to all entropies indicated above for suitable choices of the parameters.

The generalized entropy function (1) has the following properties:
1. The entropy $S_g(\alpha_+, \alpha_-, \beta, \gamma)$ satisfies the generalized third law of the thermodynamics: $S_g \to 0$, when $S \to 0$.
2. The entropy $S_g(\alpha_+, \alpha_-, \beta, \gamma)$ is a monotonically increasing function in the variable $S$, because both the terms $\left(1 + \frac{\alpha_+}{\beta} S\right)^{\beta}$ and $\left(1 + \frac{\alpha_-}{\beta} S\right)^{-\beta}$ in the expression of $S_g$ increase with $S$.
3. The entropy $S_g(\alpha_+, \alpha_-, \beta, \gamma)$ reduces to the Bekenstein-Hawking entropy at a certain limit of parameters.

For example, when $\alpha_+ \to \infty$, $\alpha_- = 0$, $\gamma = \left(\frac{\alpha_+}{\beta}\right)^{\beta}$ and $\beta = 1$ the generalized entropy function (1) becomes equivalent to the Bekenstein-Hawking entropy.

It is suggested, regarding the generalized entropy function, that the minimum number of parameters that can include all known entropies, is four [1].

**IV. Viscous entropic cosmology coupled with dark matter**

Let us apply the thermodynamic approach to describe the evolution of the universe filled with two interacting fluids: dark energy and dark matter from the four parameter generalized entropy function $S_g$, defined in [1] in a spatially flat (FRW) metric

$$ds^2 = -dt^2 + a^2(t) \sum_{i=1,2,3} (dx^i)^2. \qquad (7)$$

Here $a(t)$ is a scale factor.
The modified Friedmann equation corresponding to the generalized entropy takes the form

$$H^2 = \frac{k^2}{3}(\rho + \rho_m + \rho_g), \qquad (8)$$

where $H = \frac{\dot{a}(t)}{a(t)}$ is the Hubble function, $k^2 = 8\pi G$ is Einstein's gravitational constant with Newton's gravitational constant $G$; $p, \rho$ and $p_m, \rho_m$ are the pressure and the energy density of dark energy and dark matter, respectively. A dot denotes derivative with respect to the cosmic time $t$. Here $\rho_g$ and $p_g$ denote the energy density and pressure corresponding to the entropy function $S_g$.

Next, we consider cosmological applications of the modified Friedmann equation (8). We write the dynamic equations in the form [25]

$$\begin{cases} \dot{\rho}+3H(p+\rho)=-Q \\ \dot{\rho}_m+3H(p_m+\rho_m)=Q \\ \dot{H}=-\dfrac{k^2}{2}(p+\rho+p_m+\rho_m) \end{cases} \quad , \tag{9}$$

where $Q$ is the interaction term between dark energy and dark matter.

Let us apply the generalized entropy function $S_g$ to study of cosmological models with logarithmic (EoS). In the late-time universe, the parameter $\gamma$ becomes constant $\gamma=\gamma_0$. As a result the entropy function takes the following form

$$S_g = \frac{1}{\gamma_0}\left[\left(1+\frac{\alpha_+}{\beta}S\right)^{\beta} - \left(1+\frac{\alpha_-}{\beta}S\right)^{-\beta}\right] \tag{10}$$

with $S=\pi/(GH)^2$.

At a late stage of the evolution of the universe, the estimate of the Hubble function is of the order of $10^{-40}\,GeV$ and the condition $GH^2 \ll 1$ is satisfied. Then the formula for the energy density $\rho_g$ corresponding to the entropy function (6) can be approximated as [1]:

$$\rho_g = \frac{3H^2}{k^2}\left[1-\frac{\alpha_+}{\gamma_0(2-\beta)}\left(\frac{GH^2\beta}{\pi\alpha_+}\right)^{1-\beta}\right]. \tag{11}$$

Next, we apply the presented mathematical formalism to various types of interaction of dark energy with dark matter and different fluid viscosity forms.

### V. Entropic holographic cosmology with logarithmic-corrected viscous fluid

Let us consider a universe filled with the log-corrected power-law fluid in the presence of viscosity coupling with dark matter. Let us assume that the dark matter is dust matter, then $p_m=0$. In this case, the gravitational equation of motion of dark matter is simplified

$$\dot{\rho}_m+3H\rho_m=Q. \tag{12}$$

Next, we will consider cosmological models with the following types of interaction term $Q$.

### 1. Model with coupling term $Q=\delta H\rho_m$

For mathematical reasons, we choose the coupling term in the same form as in [27]

$$Q=\delta H\rho_m, \tag{13}$$

where $\delta$ is a positive no dimensional constant.

Let us consider the energy density $\rho_g$ in the form (11) in the case, then with $\beta=1$

$$\rho_g = \frac{3H^2}{k^2}\left(1-\frac{\alpha_+}{\gamma_0}\right). \tag{14}$$

The Friedmann equation (4) takes the form

$$H^2 = \frac{\gamma_0 k^2}{3\alpha_+}(\rho + \rho_m). \tag{15}$$

Currently, the problem of coincidences in the standard $\Lambda$CDM model remains one of the unsolved problems of modern cosmology. Since the dark energy density and the matter energy density in the modern Universe are of the same order, it can be assumed that dark energy and matter somehow interact with each other. Accurate astronomical observations show that the ratio $r = \frac{\rho_m}{\rho}$ of the order of unity. The assumption that this relation is fixed is a consequence $\Lambda$CDM models. As a consequence, equation (15) is rewritten as follows

$$\rho = \frac{3\alpha_+ H^2}{(1+r)\gamma_0 k^2}. \tag{16}$$

Using equations (14), (16) we write the interaction term $Q$

$$Q = 3\frac{(r-1)}{(r+1)}\frac{\alpha_+ \omega_0}{\gamma_0 k^2} H^3 = 3\omega_0(r-1)\rho H. \tag{17}$$

Let us choose the bulk viscosity in (4) to be proportional to the Hubble function

$$\zeta(H,t) = 3\tau H, \tag{18}$$

where the parameter $\tau$ is positive.

Let us suppose that $\rho > \frac{1}{2}\rho_*$ (density of the log-corrected power-law fluid is higher than the Planck density), and let us study the case when $l = -1$.

Using (4) the first equation from (9), (17) and (18), we obtain the gravitational equation for the viscous log-corrected power-law fluid in this approximation

$$\dot{H} - \frac{1}{2}\left[3\omega_0(1-r) + 3\frac{\tau}{\gamma} - \frac{A}{\rho_*}\right]H^2 = \frac{1}{2}\frac{A}{\gamma}, \tag{19}$$

where $\gamma = \frac{\alpha_+}{\gamma_0(1+r)k^2}$.

Let us rewrite Eq. (19) in the form

$$2\dot{H} = \tilde{a}H^2 + b, \tag{20}$$

where $\tilde{a} = 3\omega_0(1-r) + 3\frac{\tau}{\gamma} - \frac{A}{\rho_*}$ and $b = \frac{A}{2\gamma}$.

The general solution of Eq. (20) is

$$H = \sqrt{\frac{b}{a}}\left|tg\left(\frac{1}{2}\sqrt{\tilde{a}b}\,t + \sqrt{\frac{b}{\tilde{a}}}C\right)\right|, \tag{21}$$

where $C$ is an arbitrary constant.

Let's consider the particular solution (17) at $C = 0$.

In the limit $t \to \frac{1+2k}{\sqrt{\tilde{a}b}}\pi$, $k \in Z$ the Hubble function approaches infinity, $H \to \infty$, and we have quasi-periodical appearances of cosmological singularities of the Big Rip type.

For the scale factor, we obtain

$$a(t) = a_0 \exp\left[\int H(t)dt\right] = \frac{a_0}{\sqrt[\tilde{a}]{\cos^2\left(\frac{1}{2}\sqrt{\tilde{a}b}\,t\right)}}. \tag{22}$$

Next, suppose, that $A = \rho_* \left\{ 3\left[\omega_0(1-r) + \dfrac{\tau}{\gamma}\right] - 2 \right\}$, then $\tilde{a} = 2$. We obtain

$$a(t) = \dfrac{a_0}{\left|\cos\dfrac{\sqrt{2b}}{2}t\right|}.$$

(23)

The result obtained in (23) describes the fluctuations of a scale factor from 1 to infinity. This solution is not physical. It is a consequence of the choice of the form of interaction between dark energy and dark matter.

Let' us now present the results obtained in holographic form.

At first, we present the basic positions of the holographic principle in the terminology proposed by Li in [64]. Following the holographic principle, all physical quantities in the Universe can be described by certain values at the boundary of space-time [65].

In a holographic description, all physical quantities are represented through the horizon cut-off radius. According to the generalized holographic model of Nojiri and Odintsov introduced in [2], the holographic energy density is inversely proportional to the squared infrared cut-off $L_{IR}$

$$\rho_{hol} = 3c^2 k^2 L_{IR}^{-2},$$  (24)

where $k^2 = 8\pi G$ is Einstein's gravitational constant with Newton's gravitational constant $G$ and $c$ is a no dimensional in geometric units and positive constant. With this description of dark energy, the horizon cut-off radius means the infrared cut-off.

The infrared cut-off radius, describing the accelerated expansion of the universe, can be identified with the size of the particle horizon $L_p$ or the size of the event horizon $L_f$, which are defined, respectively, as

$$L_p(t) \equiv a(t) \int_0^t \dfrac{dt'}{a(t')}, \quad L_f(t) \equiv a(t) \int_t^\infty \dfrac{dt'}{a(t')},$$  (25)

where a(t) is a scale factor.

However, not every choice of infrared radius can lead to an accelerated expansion of the universe. Thus, the choice of the infrared cut-off radius is not arbitrary.

We will describe the holographic picture in terms of the particle horizon $L_p$. The Hubble function can be expressed in terms of the particles horizon and its derivatives as follows [3]:

$$H = \dfrac{\dot{L}_p - 1}{L_p}, \quad \dot{H} = \dfrac{\ddot{L}_p}{L_p} - \dfrac{\dot{L}_p^2}{L_p^2} + \dfrac{\dot{L}_p}{L_p^2}.$$  (26)

Let us calculate the particle horizon $L_p$

$$L_p = a \int_0^t \dfrac{dt}{a} = \dfrac{1}{\cos\left(\dfrac{1}{2}\sqrt{2bt}\right)} \int_0^t \cos\left(\dfrac{1}{2}\sqrt{2bt}\right) dt = \sqrt{\dfrac{2}{b}} tg\left(\sqrt{\dfrac{b}{2}}t\right) = \dfrac{2}{b}H,$$  (27)

where $b = A\dfrac{\gamma_0}{\alpha_+}(1+r)k^2$.

Further, we will assume that the fluid driving the evolution of the universe has a holographic origin.

Using (27), we can rewrite the energy conservation law (19) in the holographic form

$$\frac{\ddot{L}_p}{L_p} - \frac{\dot{L}_p^2}{L_p^2} + \frac{\dot{L}_p}{L_p^2} - \frac{1}{2}\left\{3\left[\omega_0(1-r) + \frac{\tau\gamma_0(1+r)k^2}{\alpha_+}\right] - \frac{A}{\rho_*}\right\}\left(\frac{\dot{L}_p - 1}{L_p}\right)^2 = \frac{1}{2}\frac{A\gamma_0(1+r)k^2}{\alpha_+}. \qquad (28)$$

Let us take $\alpha_+ = \gamma_0$, then the entropic energy density is $\rho_g = 0$, and the equation (28) simplifies to

$$\frac{\ddot{L}_p}{L_p} - \frac{\dot{L}_p^2}{L_p^2} + \frac{\dot{L}_p}{L_p^2} - \frac{1}{2}\left\{3\left[\omega_0(1-r) + \tau(1+r)k^2\right] - \frac{A}{\rho_*}\right\}\left(\frac{\dot{L}_p - 1}{L_p}\right)^2 = \frac{1}{2}A(1+r)k^2. \qquad (29)$$

We go thus beyond entropic cosmology.

Thereby, we have obtained a reconstruction of the energy conservation equation as a generalized form of holographic dark energy.

Let us explore the dynamics of the dark matter density, and find a solution to the gravitational equation for dark matter (12) for the coupling model (17)

$$\dot{\rho}_m + 3H\rho_m = 3\omega_0(r-1)\rho H. \qquad (30)$$

Taking into account ration $r = \frac{\rho_m}{\rho}$, we transform it to the form

$$\dot{\rho}_m = 3\left[\omega_0\left(1 - \frac{1}{r}\right) - 1\right]H\rho_m. \qquad (31)$$

Using the Hubble function (21), we obtain

$$\frac{d\rho_m}{\rho_m} = \sqrt{\frac{\tilde{a}}{b}}\theta tg\left(\frac{1}{2}\sqrt{\tilde{a}b}t\right), \qquad (32)$$

where $\theta = 3\left[\omega_0\left(1 - \frac{1}{r}\right) - 1\right]$.

The solution (27) has the form

$$\rho_m(t) = \frac{\rho_m^{(0)}}{\left|\cos\left(\frac{1}{2}\sqrt{\tilde{a}b}t\right)\right|^{\frac{2\theta}{\tilde{a}}}}. \qquad (33)$$

If $t \to \frac{2k}{\sqrt{\tilde{a}b}}\pi$, $k \in Z$ then the initial value becomes $\rho_m(t) \to \rho_m^{(0)}$. After that, in the limit $t \to \frac{1}{\sqrt{\tilde{a}b}}(k+1)\pi$, the energy density increases infinitely $\rho_m(t) \to \infty$. Then, the density of dark matter decreases and reaches the initial value $\rho_m^{(0)}$, after which it increases again infinitely.

Thus, the physical consequence of solving equation (33) is a periodic change in the density of dark matter.

## 2. Model with coupling term $Q = 3\lambda H \frac{\rho\rho_m}{\rho + \rho_m}$

Next consider the interacting model with coupling term in the form

$$Q = 3\lambda H \frac{\rho\rho_m}{\rho + \rho_m}, \qquad (34)$$

where the parameter $\lambda$ is a dimensionless constant.

In terms of the energy density ratio $r = \frac{\rho_m}{\rho}$ the interaction term can be written

$$Q = \frac{3\lambda r}{1+r}H\rho. \qquad (35)$$

Then the gravitational equation for the log-corrected power-law coupled fluid in the presence of the constant bulk viscosity $\zeta(H,t) = \zeta_0$ in the approximation $\rho > \dfrac{\rho_*}{2}$ for the case $l = -1$ will take the form

$$\dot{H} + \frac{1}{2}\left(\frac{A}{\rho_*} - \frac{\lambda r}{1+r} + 1\right)H^2 - \frac{3\zeta_0}{2\gamma}H = \frac{1}{2}\frac{A}{\gamma}, \qquad (36)$$

where $\gamma = \dfrac{\alpha_+}{\gamma_0(1+r)k^2}$, as before.

Let us introduce the notation $a_1 = \dfrac{1}{2}\left(\dfrac{A}{\rho_*} - \dfrac{\lambda r}{1+r} + 1\right)$, $a_2 = \dfrac{3\zeta_0}{2\gamma}$ and $a_3 = \dfrac{1}{2}\dfrac{A}{\gamma}$, whereby we transform this equation to the form

$$\frac{dH}{a_1 H^2 - a_2 H - a_3} = -dt. \qquad (37)$$

Let us consider the case when the determinant of the quadratic equation $a_1 H^2 - a_2 H - a_3 = 0$ is equal zero, that is $a_2^2 + 4a_1 a_3 = 0$.

From this follows the equation

$$A^2 + \left(1 - \frac{\lambda r}{1+r}\right)A\rho_* + (1+r)\frac{\gamma_0 \rho_*}{\alpha_+}\left(\frac{3}{2}\zeta_0 k\right)^2 = 0. \qquad (38)$$

The solution (33) has the form

$$A_{1,2} = -\frac{1}{2}\left(1 - \frac{\lambda r}{1+r}\right)\rho_* \pm \sqrt{\frac{1}{4}\left(1 - \frac{\lambda r}{1+r}\right)^2 \rho_*^2 - 4(1+r)\frac{\gamma_0 \rho_*}{\alpha_+}\left(\frac{3}{2}\zeta_0 k\right)^2}. \qquad (39)$$

Let us suppose that $\alpha_+ = \dfrac{(1+r)\gamma_0}{\rho_*}\left(\dfrac{6\zeta_0 k}{\dfrac{\lambda r}{1+r} - 1}\right)^2$, then the radical expression (34) vanishes and the solution of the equation (33) simplifies: $A = \dfrac{1}{2}\left(\dfrac{\lambda r}{1+r} - 1\right)\rho_*$.

Then we can write the Hubble function in the form

$$H = \frac{1}{a_1 t + C_1} + \frac{a_2}{2a_1} = \frac{1}{\frac{1}{4}\left(1 - \frac{\lambda r}{1+r}\right)t + C_1} + \frac{\gamma_0 \rho_*}{12\zeta_0}, \quad (40)$$

where $C_1$ is an arbitrary constant.

If $t$ tends to the finite value $t \to -\dfrac{C_1}{\frac{1}{4}\left(1 - \frac{\lambda r}{1+r}\right)}$, then $H \to \infty$, it take place the cosmological singularity of the Big Rip type.

In the limit $t \to \infty$ the Hubble function tends to the cosmological constant $H \to \dfrac{\gamma_0 \rho_*}{12\zeta_0}$.

Next, let us move on to the holographic description in this model of the interaction of dark energy and dark matter.

Let us calculate the scale factor,

$$a(t) = a_0 \exp\left[\int H(t)dt\right] = a_0 \left[\frac{1}{4}\left(1 - \frac{\lambda r}{1+r}\right)t + C_1\right]^{\frac{4}{1 - \frac{\lambda r}{1+r}}} e^{\frac{\gamma_0 \rho_*}{3\zeta_0\left(1 - \frac{\lambda r}{1+r}\right)}t}, \quad (41)$$

In the absence of interaction between dark energy and dark matter, when the parameter $\lambda = 0$, we obtain

$$a(t) = a_0 \left(\frac{1}{4}t + C_1\right)^4 e^{\frac{\gamma_0 \rho_*}{3\zeta_0}t}. \quad (42)$$

Let us calculate, using (37), the particle horizon $L_p$ in the case when $a_1 = 1$. It then follows that the interaction parameter is equal to $\lambda = -3\left(1 + \dfrac{1}{r}\right)$, and

$$L_p = a \int_0^t \frac{dt}{a} = (t + C_1) e^{\frac{\gamma_0 \rho_*}{12\zeta_0}t} \int_0^t \frac{e^{-\frac{\gamma_0 \rho_*}{12\zeta_0}t}}{t + C_1} dt = (t + C_1) e^{\frac{\gamma_0 \rho_*}{12\zeta_0}(t - C_1)} Ei\left(-\frac{\gamma_0 \rho_*}{12\zeta_0}C_1 - \frac{\gamma_0 \rho_*}{12\zeta_0}t\right), \quad (43)$$

where $Ei(x)$ is the integral exponential function and $C_1$ is an arbitrary constant.

We see, using (26), that in this approximation the corresponding holographic representation of the energy conservation equation (36) takes the form

$$\frac{\ddot{L}_p}{L_p} - \frac{\dot{L}_p^2}{L_p^2} + \frac{\dot{L}_p}{L_p^2} - \left(\frac{\dot{L}_p - 1}{L_p}\right)^2 + \frac{1}{12}\frac{\rho_*}{\zeta_0}\left(\frac{\dot{L}_p - 1}{L_p}\right) = \frac{1}{3}\left(\frac{\rho_*}{\zeta_0}\right)^2. \tag{44}$$

Thus, we have obtained a holographic representation of the logarithmically corrected power-law model for a viscous dark fluid coupled with dark matter.

Next, we consider the gravitational equation of motion of dark matter (12) with coupling term in the form (34).

$$\dot{\rho}_m + 3H\rho_m = 3\lambda H \frac{\rho\rho_m}{\rho + \rho_m}. \tag{45}$$

Let us rewrite it as follows

$$\dot{\rho}_m + 3\left(1 - \frac{3\lambda}{1+r}\right)H\rho_m = 0. \tag{46}$$

Using the Hubble function (35), we obtain

$$\frac{d\rho_m}{\rho_m} = -3\left(1 - \frac{3\lambda}{1+r}\right)\left(\frac{1}{\frac{1}{4}\left(1 - \frac{\lambda r}{1+r}\right)t + C_1} + \frac{\gamma_0 \rho_*}{12\zeta_0}\right)dt. \tag{47}$$

Integrating, we obtain a solution (42) in the form

$$\rho_m(t) = \rho_m^{(0)}\left(a_1 t + C_1\right)^{\frac{b_1}{a_1}} e^{a_2 b_1 t}, \tag{48}$$

where $a_1 = \frac{1}{4}\left(1 - \frac{\lambda r}{1+r}\right)$, $a_2 = \frac{\gamma_0 \rho_*}{12\zeta_0}$, $b_1 = 3\left(\frac{3\lambda}{1+r} - 1\right)$ and $\rho_m^{(0)}$ is a constant of integration.

If $a_1 = 1$, then $\lambda < 0$ and the parameter $b_1$ is negative. Since the parameter $a_2$ is positive, then in the far future (in the limit $t \to \infty$) the density of dark matter tends to zero.

### 3. Model with coupling term $Q = 3\lambda H(\rho + \rho_m) + \delta(\dot{\rho} + \dot{\rho}_m)$

Let us assume that the coupling between dark energy and dark matter has the form

$$Q = 3\lambda H(\rho + \rho_m) + \delta(\dot{\rho} + \dot{\rho}_m), \tag{49}$$

where the parameters $\lambda$ and $\delta$ are dimensionless constants. Thus, the interaction between the fluid components depends on the sign of $\lambda$ and $\delta$. This assumption is related to that in [66].

We choose the viscosity to be proportional to the Hubble function

$$\zeta(H, t) = 3\tau H, \tag{50}$$

where the parameter $\tau$ is positive.

Taking into account the assumptions made above, in the approximation $\rho > \frac{\rho_*}{2}$ for the case $l = -1$ the energy conservation law for the dark fluid takes the form

$$\frac{2\alpha_+}{\gamma_0 k^2}\left(\frac{1}{1+r} + \delta\right)\dot{H} + \left\{\frac{3\alpha_+}{\gamma_0(1+r)k^2}\left[\frac{A}{\rho_*} + \lambda(1+r) + 1\right] - 9\tau\right\}H^2 = A. \tag{51}$$

Let us denote $\tilde{a} = \frac{2\alpha_+}{\gamma_0 k^2}\left(\frac{1}{1+r} + \delta\right)$, $\tilde{b} = \frac{3\alpha_+}{\gamma_0(1+r)k^2}\left[\frac{A}{\rho_*} + \lambda(1+r) + 1\right] - 9\tau$ and rewrite the equation (51) in the form

$$\tilde{a}\dot{H} + \tilde{b}H^2 = A. \tag{52}$$

The Hubble function will take the form

$$H = \sqrt{\frac{A}{\tilde{b}}} \frac{C_2 e^{2\frac{\sqrt{A\tilde{b}}}{\tilde{a}}t} - 1}{C_2 e^{2\frac{\sqrt{A\tilde{b}}}{\tilde{a}}t} + 1}, \tag{53}$$

where $C_2$ is an arbitrary constant.

Let us take $C_2 = 1$, then in the limit $t \to \infty$ the Hubble function tends to the cosmological constant $H \to \sqrt{\frac{A}{\tilde{b}}}$.

Let us calculate the scale factor

$$a(t) = a_0 e^{\int H(t)dt} = a_0 \left( 2ch\frac{2\sqrt{A\tilde{b}}}{\tilde{a}}t \right)^{\frac{\tilde{a}}{\tilde{b}}}. \tag{54}$$

From the condition $A = \rho_* \left[ (1+r)\left( \frac{\delta}{3} - \lambda + \frac{3\tau\gamma_0 k^2}{\alpha_+} \right) - \frac{2}{3} \right]$ it follows that $\tilde{a} = \tilde{b}$.

Taking this equality into account, we calculate the particle horizon $L_p$.

$$L_p = a\int_0^t \frac{dt}{a} = ch\left(2\sqrt{\frac{A}{\tilde{a}}}t\right)\int_0^t \frac{dt}{ch\left(2\sqrt{\frac{A}{\tilde{a}}}t\right)} = 2\sqrt{\frac{A}{\tilde{a}}}\left[ arctg\, e^{2\sqrt{\frac{A}{\tilde{a}}}t} ch\left(2\sqrt{\frac{A}{\tilde{a}}}t\right) - \frac{\pi}{4} \right]. \tag{55}$$

Let us present the energy conservation law (51) obtained in holographic language

$$\frac{2\alpha_+}{\gamma_0 k^2}\left( \frac{1}{1+r} + \delta \right)\dot{H} + \left\{ \frac{3\alpha_+}{\gamma_0(1+r)k^2}\left[ \frac{A}{\rho_*} + \lambda(1+r) + 1 \right] - 9\tau \right\}H^2 = A. \tag{56}$$

Next, let us consider the change in the density of dark matter over time. Let us find a solution of the gravitational equation of motion for dark matter with the coupling term (49).

$$\dot{\rho} + 3H\rho_m = 3\lambda H(\rho + \rho_m) + \delta(\dot{\rho} + \dot{\rho}_m). \tag{57}$$

Using the ratio $r = \frac{\rho_m}{\rho}$, we can rewrite it in terms of energy density of dark matter

$$\left[ 1 - \delta\left(1 + \frac{1}{r}\right) \right]\dot{\rho}_m = 3H\rho_m\left[ \lambda\left(1 + \frac{1}{r}\right) - 1 \right]. \tag{58}$$

This leads to

$$\frac{\dot{\rho}_m}{\rho_m} = 3\theta\sqrt{\frac{A}{\tilde{a}}}\frac{e^{\sqrt{\frac{A}{\tilde{a}}}t} - 1}{e^{\sqrt{\frac{A}{\tilde{a}}}t} + 1}, \tag{59}$$

where $\theta = \frac{\lambda(1+r) - r}{r - \delta(1+r)}$.

The solution of equation (59) becomes

$$\rho_m(t) = \rho_m^{(0)} \left( 2ch2\sqrt{\frac{A}{\tilde{a}}}t \right)^{3\theta}. \tag{60}$$

In the limit $t \to \infty$ at $\theta > 0$ dark energy density increases $\rho_m(t) \to \infty$, and at $\theta < 0$ tends to zero.

Thus, in the entropic cosmology in the far future it is possible that the dark matter energy density increases increase indefinitely. And yet there is the possibility that it can simply disappear.

## VI. Conclusion

In this work, we studied, based on a thermodynamic approach, a model of dark energy with a modified logarithmic equation of state, taking into account the interaction of dark energy with dark matter. It was assumed that the dark fluid possessed viscosity.

Homogeneous and isotropic Friedman-Robertson-Walker space-time was considered. The choice of a logarithmic equation of state was due to the fact that a dark fluid with such an equation of state has properties similar to those of crystalline solids under isotropic deformations in cases where the pressure is negative.

The novelty of this work lies in the application of a thermodynamic approach based on the generalized entropy function. A physical consequence of the entropy cosmology model is the generation of entropy energy density via entropy. This leads to the inclusion in the Friedmann equation of an energy density that corresponds to a generalized entropy function. It leads also to modification of the parameters in the (EoS). Thus, the Friedmann equation becomes a consequence of the fundamental laws of thermodynamics.

Different types of interactions between dark energy and dark matter were studied. Various forms of bulk viscosity in the logarithmic equation of state of a dark fluid were considered. Taking into account the entropy energy density in the Friedmann equation, solutions to the gravitational equation of motion for a dark fluid were obtained. Analysis of solutions showed that taking into account the thermodynamic properties of the system leads to a singular behavior of the Hubble function of the Big Rip type. It is possible that the Hubble function tends asymptotically to the cosmological constant.

The gravitational equations of dark matter were solved. Analysis of the solutions showed that in the far future, at certain values of the parameters, a scenario is possible in which the density of dark matter will increase. A quasi-periodic change in the dark matter density function is possible. It can be assumed that in the far future the role of dark matter in the process of the evolution of the Universe will change.

We presented the gravitational equations for dark energy density in holographic form. In order to apply the holographic principle to these models, we identified the infrared radius $L_{IR}$ with the particle horizon $L_p$. In these models, the infrared radiuses in the form of a particle horizon were calculated, in order to obtain the appropriate energy conservation law in each case. As a result, we showed the equivalence between the viscous fluid models and the holographic models.

Research into the evolution of the late and early Universe based on the new generalized entropy was also carried out in the works [67, 68].

It would be interesting to extend this formulation with account of thermal effects [69].